\documentclass[a4paper,12pt]{article}

\usepackage{amstext,amsmath,amssymb,amsthm}  

\usepackage[normalem]{ulem}  
\usepackage{graphicx}  
\usepackage{bm} 

\usepackage[comma,sort,authoryear]{natbib}
\usepackage{hypernat}  

\usepackage{vmargin}  
\usepackage{authblk}

\usepackage{xcolor}
\definecolor{dark-red}{rgb}{0.8,0.15,0.15}
\definecolor{dark-blue}{rgb}{0.15,0.15,0.6}
\definecolor{medium-blue}{rgb}{0,0,0.8}

\usepackage[pdftex,breaklinks,colorlinks]{hyperref}
\hypersetup{  
  pdftitle =
    {The relevance of ontological commitments},
  linkcolor={dark-red},
  citecolor={dark-blue},
  urlcolor={medium-blue}
}

\begin{document}

\numberwithin{equation}{section}
\numberwithin{figure}{section}
\allowdisplaybreaks[1]  

\title{The relevance of ontological commitments}

\author[,1,2,3,4,5]{Pablo Echenique-Robba\footnote{{\footnotesize \
\href{mailto:pablo.echenique.robba@gmail.com}{\texttt{pablo.echenique.robba@gmail.com}} ---
\href{http://www.pabloecheniquerobba.com}{\texttt{http://www.pabloecheniquerobba.com}}}}}

\affil[1]{Instituto de Qu\'{\i}mica F\'{\i}sica Rocasolano, CSIC, Madrid, Spain}
\affil[2]{Instituto de Biocomputaci\'on y F{\'{\i}}sica de Sistemas Complejos (BIFI), Universidad de Zaragoza, Spain}
\affil[3]{Zaragoza Scientific Center for Advanced Modeling (ZCAM), Universidad de Zaragoza, Spain}
\affil[4]{Departamento de F{\'{\i}}sica Te\'orica, Universidad de Zaragoza, Spain}
\affil[5]{Unidad Asociada IQFR-BIFI, Madrid-Zaragoza, Spain}

\date{\today}

\maketitle

\begin{abstract}

In this introductory note, I describe my particular view of the notion of
\emph{ontological commitments} as honest and pragmatic working hypotheses that
assume the existence (out there) of certain entities represented by the symbols
in our theory. I argue that this is not naive, in the sense that it does not
entail the belief that the hypotheses could ever be proved to be true (or
false), but it is nevertheless \emph{justified} by the success and predictive
power of the theory that contains the concepts assumed to exist. I also claim
that the ontological commitments one holds (even if tacitly so) have a great
influence on what kind of science is produced, how it is used, and how it is
understood. Not only I justify this claim, but I also propose a sketch of a
possible falsification of it. As a natural conclusion, I defend the importance
of identifying, clarifying and making explicit one's ontological commitments if
fruitful scientific discussions are to be had. Finally, I compare my point of
view with that of some philosophers and scientists who have put forward similar
notions.
\vspace{0.4cm}
\vspace{0.2cm}\\

\end{abstract}

\section{Introduction}
\label{sec:introduction}

In some physics textbooks the importance of ontology (i.e., what is assumed to
exist) is strongly emphasized \citep{Durr2009}, while in some others ontology is
banished out of physics \citep{Peres2002}. Some physicists declare themselves
fervent believers in this or that ontology, while some others allege complete
agnosticism and call it a virtue.

In this note, I discuss a middle point which is closer to the first kind of
positions, but only if properly qualified and not held in a naive way. Although
I am certainly not introducing any new idea, I hope that the personal nuances
that configure my particular view of the issue can serve the reader to see it
in a new light, and perhaps open a debate which I consider very necessary.

\section{My view of the concept}
\label{sec:my_view}

\subsection{Ontological certainty vs. ontological commitment}
\label{subsec:certainty_vs_commitment}

Our access to reality is, in general, \emph{indirect}. That much is obvious.
Hence, it seems very difficult that we could ever be \emph{completely sure}
about the ontological status of the concepts to which our symbols refer. That
is, about whether or not what we collectively mean by, say, ``electron''
actually \emph{exists} out there, or if it simply denotes some complicated way
that we humans have found to summarize certain information that we might obtain
about an unknown underlying reality. Of course, since the ``electron'' concept
is very useful, it has enormous predictive power, and it fits nicely and
consistently with a myriad of other concepts that are also useful and
predictive, even if the answer was the latter, the information that
``electron'' summarizes must be strongly correlated with whatever exists out
there. It is in this sense that the information that we bookkeep under the
``electron'' label can be said to be \emph{about} the underlying reality; even
if it is not made of ``electrons'' at all.

Maybe this is the moment to remark that it is obvious to me that
\emph{something exists}, i.e., that it is not the case that nothing exists.
Therefore, \emph{there is} (to me) some underlying, external, objective
reality. The only questions are of the type of: What is it? Is it in principle
possible to find out what it is? Do humans have what is required to
\emph{actually} find out what it is? In this particular and very restricted
conversation, are ``electrons'' a part of it? Etcetera.

These are very difficult questions indeed, and I don't know what the correct
answers might be; nor I think that anyone does. As I said, it is conceivable to
me that ontological certainty could even be \emph{unattainable} in principle
(or in practice, which, in this case, looks very much like the same thing). If
this is so, then the actual truth value of ontological statements such as ``the
electron exists'' will be forever beyond our reach, and therefore it could be
argued that it is irrelevant. This might be so, but even in such a case what is
\emph{not} irrelevant is \emph{our commitment} to the truth of such statements;
a powerful concept that I first found in \citep{Maudlin2007}, and which now I
take the opportunity to put in my own words.

\subsection{Ontological commitments in everyday life}
\label{subsec:in_everyday_life}

Everyday life is full of concepts whose existence (out there) is very
questionable but the fact that humans are ontologically committed to it is
nevertheless incredibly important: God, democracy, free markets, beauty,
afterlife, are some examples from a very long list.

These commitments are important mainly because they guide \emph{behavior}
(individual and collective), and we can even drop the adverb ``mainly''; after
all, there is hardly anything more important than behavior for us humans. At
least in everyday life, pragmatically, and understood in a wide sense (talking,
even thinking, is behavior too; we don't have to move limbs to \emph{do}
things), it seems clear that what we do, what we say and what we think have an
enormous influence in our lives. Even if natural phenomena such as the weather,
most of our own biological processes, and the behavior of other humans and
other animals are more or less independent of what we do (and thus quite
independent of what we are ontologically committed to), we can sometimes
anticipate these external events, we can certainly react to them, and we can
even modify them in some occasions. The nature of these interactions and their
likelihood of being beneficial to us or to others is again strongly dependent
on what we believe to exist.

\subsection{And in science}
\label{subsec:in_science}

We could think that science is different from everyday life in this respect,
but we would be wrong. Despite the sometimes grandiose statements that come
from it, the almost mythological image that some of its practitioners have
acquired, and the fact that it often seems to be worried only with the most
abstract of abstractions, science is still done by scientists (for the moment),
and scientists are still humans (really). Then, my main claim should come as no
surprise: \emph{What one is ontologically committed to is extremely relevant
regarding what kind of science one produces}. I will try to justify this in what
follows, but first.

\subsection{The radical positivist complains}
\label{subsec:radical_positivist}

At this point, the radical positivist may rise his hand and say something like
``Only what is measured is important!'', or ``You can only be committed to the
existence of what you measure!'', or even ``Only quantities that can be
measured should enter the formalism of your theory!'' The debate has been
running for many centuries, and a lot of qualifications would be in order if I
were to provide a balanced and thorough account of the different positions.
However, I have no intention to do so. The radical positivist viewpoint is so
weak, so naive, that even an amateur philosopher like myself with only a 
superficial knowledge of the issues involved can easily refute it.

The basic point is that radical positivists do not speak about \emph{how things
are}, but about \emph{how they think they should be}. Perhaps the ultimate
theory of nature contains no unmeasurable concepts, but every theory we know of
today contains them. Perhaps a superior race of aliens thinks only in terms of
concepts they can measure, but humans don't. Perhaps we should strive very
strongly to eliminate everything that cannot be measured from our language and
from our scientific theories, but we seem extremely far from achieving that
feat. As things stand today, both everyday life and science are absolutely full
of unmeasurable concepts, and in both fields these concepts have proved
extremely useful.

Thus, let us simply dare our radical positivist friend to decide what kind of
toothpaste he should buy next week without using any unmeasurable
concept---such as ``next week''---and let us move on while he struggles
endlessly with the challenge.

\subsection{Measurements are not the whole story}
\label{subsec:measurements_not_all}

Measurable concepts in a theory have of course the crucial function of being
the \emph{loci} at which we can compare it to experimental data, and thus
possibly falsify it (or produce useful predictions and technologies). However,
this is by no means the end---nor the beginning---of the story.

In order to have a theory to begin with, we have to create it first. In order
to make testable predictions with it (once created), we need to know what
manipulations of the concepts in the theory are valid and which aren't, what
counts as a prediction and what as an intermediate result, what can be \emph{in
principle} observed. Sometimes we also need to know how to interface the theory
with some other part of the theoretical corpus that uses very different
concepts (e.g., quantum mechanics and thermodynamics). In order to devise a new
theory or modify the existing one, if it is in fact falsified by experiment, we
need to be able to decide---more often to guess!---what kind of equations could
be written or which ones could be tweaked, what parts of the theory look
dispensable and what parts look \emph{in}dispensable

To achieve any of these ends, we certainly have to take into account measurable
concepts (after all, agreement with experiment is compulsory), but this is not
typically \emph{enough} to clearly illuminate the correct way forward. We often
have many options---or none at all---and we need other kinds of inputs to break
the \emph{impasse}. They could come from mathematical considerations, such as
consistency; aesthetic ones, such as simplicity; from other well settled
theories; or from ontological commitments. The commitment to the existence of
atoms in the explanation of Brownian motion, or of light quanta in the
explanation of the photoelectric effect, are clear examples of the latter kind
of input.

I am not an experimentalist, but I have the strong suspicion that \emph{also}
when designing, performing and analyzing experiments, what researchers do, say
and think, what course of action is chosen, and what is published, are all very
much influenced by their ontological commitments regarding whatever piece of
the physical world is sitting on the laboratory table, or is being observed in
the field or in a distant galaxy. It seems obvious to me that experiments are
informed by theory, and I have just argued that theory \emph{is} dependent on
ontological commitments. So the conclusion follows.

\subsection{In a nutshell}
\label{subsec:nutshell}

In sum, to commit to an ontological statement such as ``electrons exist out
there and not only in my theory and in my mind'' is not equivalent to naively
asserting that we will be able (some day) to prove the statement true (or
false). On the contrary, it is a pragmatic and honest recognition that thinking
that electrons actually exist is a powerful \emph{working hypothesis}; i.e.,
that making such ontological commitment will help us write more predictive,
simpler theories with a wider range of application, or to understand and apply
the ones we already have more soundly and faster.

This is a true commitment and not just a mask that one \emph{only} wears for
pragmatic reasons because, if we happen to be right, and we \emph{are} in fact
led to better theories or we \emph{are} able to use the ones we have more
efficiently by assuming that electrons exist, then we will be justified to
think that the electron concept is closer to the ultimate underlying reality
than the old concepts the real existence of which suggested the previous, less
powerful theory. We will be justified to think that, even if we cannot prove it.

\subsection{The importance of being explicit and clear}
\label{subsec:explicit_and_clear}

If you agree with me that ontological commitments are \emph{this} important,
then you should also agree that it is advisable that they are made
explicit---and clear---in any presentation (or more or less subtle application)
of the theories that depend on them.

Unfortunately, this is rather uncommon, and many discussions are muddled
because of the omission. In fact, what brought me to the present reflections
was the realization that, in quantum mechanics---a specially muddled
field---some authors are committed to the existence of the wave function, some
prefer to think that observables are the real thing, and some others assume
that neither of them are really out there. Most of the times, these ontological
commitments are tacit, but they are nevertheless there; you can read them
between the lines and in the equations that are written explicitly. In the
worst cases, the authors are simultaneously committed to the existence of
contradictory entities (typically without knowing it), which is the perfect
breeding ground for the infamous ``paradoxes''.

To make things even more complicated, sometimes all of this is combined with
explicit statements about the necessity of \emph{not making} any ontological
commitment (which could be seen by some physicists as a sign of weakness, a
sign that you might be a covert philosopher trying to sabotage physics from the
inside). Of course, this recommendation is extremely difficult to follow and
very tiresome; much like ``don't think about a pink elephant'' or even ``don't
think about anything at all''. Making ontological commitments is just something
that our human brains \emph{do automatically}, and I apologize for the
repetition but scientists are humans, and yes, they have brains. When somebody
recommends not making ontological commitments, what they really mean is a
creative combination of (i) try very hard not to make them (even if this means
that you move very slowly), (ii) if you catch yourself making them feel ashamed
and try harder, and (iii) in any case, \emph{never ever} make them
explicit---if you cannot avoid it, at least keep it for yourself. In that way,
nobody could argue in a trial that you have in fact made such a terrible
mistake, and maybe you could even maintain (for yourself) the self-reassuring
fiction that your agnosticism is pure.

Caricatures aside, it is my opinion that clearly stating these important
ingredients (the ontological commitments which you certainly make---like it or
not---and which strongly influence how you do science) could be very useful to
put some order in the field of foundations of quantum mechanics; a field which
is teeming at the moment with tens of versions of the theory (sometimes called
``interpretations''), and with researchers talking past each other in
specialized journals and conferences \citep{Echenique-Robba2013}. However, it
is worth pointing out that the general argument in favor of the importance of
ontological commitments---and therefore of stating them explicitly and
clearly---does not apply only to quantum mechanics, but across the scientific
board.

\subsection{A sketch of a falsification}
\label{subsec:sketch_falsification}

To me, it is obvious that ontological commitments (i) are unavoidable features
of human thought, and (ii) they affect what you do, what you say and what you
think, in everyday life and in science. However, these are claims about the
psychology of human beings, not about any underlying reality. Hence, they are
in principle open to experimental falsification. You don't have to take my word
for it---or reject it. \emph{We can check who's right}.

I am no experimental psychologist, but I have already proved that my condition
of amateur doesn't deter me from irresponsibly exploring philosophy or 
experimental physics. Thus, I will continue along my foolish line and I will 
provide a very rough sketch of some experiments that may be done to actually 
falsify the two claims in the previous paragraph. Of course, these are just the
germ of the beginning of a proper brainstorm which should be conducted, ideally,
by a collaboration between philosophers, theoretical physicists and experimental
psychologists. I am only writing this up to provoke and to (hopefully) 
jumpstart a useful discussion.

\begin{itemize}

\item \textbf{Are ontological commitments unavoidable features of human
thought?} A proper falsification of the claim would maybe amount to finding
\emph{at least one human} who \emph{at least in a given field of knowledge} can
be shown not to make any ontological commitments. It seems clear that we make
them all the time in everyday life, so we can't look there. To begin our
search, we can take some physicist who declares himself a complete agnostic,
say, about what the ontological elements in his view of quantum mechanics are.
If he is right, he constitutes a counterexample. However, his mere statement
that this is the case is definitely not enough; humans are famous for not
knowing themselves very well. Maybe we should look in his conscious behavior
for clues that signal the existence of unconscious ontological commitments---or
the absence of them. Perhaps we could look in the language he uses when he
talks informally about the theory, or in his reaction to carefully crafted
sentences that (subtly) imply an ontological commitment as compared to others
that don't. It doesn't seem easy to do it in a careful, controlled and
convincing way, but it doesn't seem impossible either.

\item \textbf{Do ontological commitments affect what you do, what you say and
what you think when producing or applying science?} This is easier to check, at
least if one is willing to use students as guinea pigs. Take two groups of
students of similar level, attitude and proficiency but who have never been
formally exposed to quantum mechanics. Teach one of them one ``interpretation''
of quantum mechanics with a given ontological import [say, Bohmian mechanics
\citep{Durr2009}], teach the other a different ``interpretation'' [say, some
version of instrumentalism such as the one in \citep{Peres2002}]. Then, ask the
two groups the same set of questions about quantum mechanics, explain to them
some subtle result (such as Bell's inequalities), propose to them a number of
exercises, and measure the quality of the answers, the depth of their
understanding and the ability to solve problems. Finally, compare the two sets
of results and see whether they are significantly different.

You might argue that Bohmian mechanics and Peres' instrumentalism are not only
distinct at the ontological level, but also at the formal one (which could
contaminate the conclusions). You could also object that \emph{neither of them}
requires to make any specific ontological commitments, i.e., that you can
commit to anyone of them while remaining agnostic about ontology. I might
answer to both objections at the same time by proposing that you feed \emph{the
same} formalism to both groups, adding, for each one of them \emph{different
and explicit} ontological viewpoints. However, as I said, my aim here is just
to open the door to a possible falsification of my claims, not to fix every
detail of the experiments. So I will stop the analysis here and move on.

\end{itemize}

\section{Others' views}
\label{subsec:others}

As with any philosophical concept, that of \emph{ontological commitment} can
take many different forms and depths; even if they are, of course, very much
related. The version that I have just presented here also has its idiosyncratic
characteristics; mainly stemming from the fact that it is \emph{my} version: it
is rather naive and superficial from the philosophical point of view, it is
eminently pragmatic, and it is focused on the business of doing science
(specifically, physics).

My notion is very similar to the version that Sean Carroll briefly mentions in
this blog post \citep{Carroll2012}, and it also has much in common with Tim
Maudlin's view, which I have already mentioned \citep{Maudlin2007}. I am sure
that Carroll's version is not as naive as mine, but since his post is really
short, we cannot tell. Maudlin's version, on the other hand, is certainly not
naive or superficial, and we \emph{can} tell by the length and the depth of the
analysis.

In chap.~3 of \citep{Maudlin2007}, we can find a cogent discussion about
ontology that includes the concept of ontological commitment as traditionally
conceived in philosophy, i.e., as the question about what is the ontology that
a given theory commits us to if we accept the latter. Maudlin discusses the
celebrated proposal by Quine and also further developments, and he concludes
that they are insufficient. The main claim by Maudlin is that:

\begin{quote}
{\small \ldots metaphysics, i.e. ontology, is the most generic account of what 
exists, and since our knowledge of what exists in the physical world rests on 
empirical evidence, metaphysics must be informed by empirical science.}
\end{quote}

In particular, and as he exemplifies later, it must be informed by the best
physical theories we have. The prevalence that Maudlin assigns to physical 
theories over philosophical speculation is clearly stressed in the following
paragraph at the end of the same chap.~3:

\begin{quote}
{\small Fiber bundles provide new mathematical structures for representing 
physical states, and hence a new way to understand physical ontology. For 
example, modern electromagnetic theory holds that what we call the 
`electromagnetic field' just \emph{is} the connection on a fiber bundle. Such 
an account evidently carries with it quite a lot of ontological commitments: 
there must be a base space, and internal degrees of freedom at each base point 
represented by a fiber, and a unified object that corresponds to all of the 
fibers with a connection. But if one asks whether, in this picture, the 
electromagnetic field is a \emph{substance} or an instance of a 
\emph{universal} or a \emph{trope}, or some combination of these, none of the 
options seems very useful [something which he proved in the previous sections]. 
If the electromagnetic field is a connection on a fiber bundle, then one 
understands what it is by studying fiber bundles directly, not by trying to 
translate modern mathematics into archaic philosophical terminology. If an 
electromagnetic field is a connection on a fiber bundle, then there are more 
things in heaven and earth than dreamt of in Plato's or Aristotle's philosophy. 
And surely this is to be expected: it would be a miracle if the fundamental 
ontological structure of the universe fit neatly in the categorical pigeonholes 
handed down to us from two millennia ago.}
\end{quote}

In my naive account of ontological commitment, I have also focused on science
in general and physics in particular, and I agree with Maudlin that scientific
theories are one of the most important sources of ontological insight. However,
my words also suggest that I also count \emph{intuition} among my
sources---which I do. Although this might seem to run counter to the previous
criticism by Maudlin against philosophical speculation, this impression is
short lived. As I see it, the most important thing you must do in order to
arrive to an ontology that might be useful (and thus possibly closer to the
ultimate underlying reality) is indeed to look at the best scientific theories
we have; exactly as Maudlin claims. However, this is (i) difficult and (ii)
maybe not enough. I.e., when you are confronted with a theory (say, quantum
mechanics), you might conceive of several different ways of inferring an
ontology from it---or none at all. In fact, this seems to be exactly what is
happening nowadays in the field. This suggests, as I say, that the program of
deriving your ontological commitment from scientific theories is sensible, but
also very difficult and perhaps underdetermined. Exactly the kind of problem
where good, informed intuition can become handy.

As Hilary Putnam puts it in a nice aphoristic way \citep{Putnam2012}:

\begin{quote}
{\small \ldots mathematically presented quantum mechanical theories do not wear 
their ontologies on their sleeve.}
\end{quote}

Judging by Maudlin's critique of Quine's approach, which emphasizes that some
metaphysical heavy lifting is involved in translating a theory into a form in
which it can be used to infer ontological commitments, as well as by the fact
that he outlines the program but does not take it to an end---what an amazing
feat would that be!---I think that he might agree with me (and with Putnam) on 
this qualification.

Also much related to this point is of course the fact that our different
physical theories present distinct levels of formalization and conceptual
tidiness. Some are compact, beautifully axiomatized and crystal clear about
what they claim and how they work, but some others are not quite so. The task
of extracting ontological commitments from the latter family of theories is
probably more difficult than doing the same for the former, and intuition plays
a more determinant role in such a case too.

Another philosopher who famously informs his philosophy on science is Daniel
C.~Dennett. In fact, we can also find in Dennett's account of the
\emph{intentional stance} an idea which is so similar to my
\emph{justification} of ontological commitments through the power of the theory
that rests on them that I have strong doubts about whether the idea is mine at
all, or I simply got it from him and then forgot its origin. In any case, since
Dennett puts it so clearly, and the reasoning also serves as a bridge between
Maudlin's theories prevalence and my moderate intuitionism, let me briefly
discuss the issue here.

As we can learn in sec.~18 of the introductory book \citep{Dennett2013}, and
also in the much more detailed \citep{Dennett1989}, the ``intentional stance''
is a level of abstraction, a mode of reasoning if you will, which is specially
good in dealing with certain kind of systems termed (of course) ``intentional
systems'' or sometimes just ``agents''. In his own words \citep{Dennett1989}:

\begin{quote}
{\small Here is how it works: first you decide to treat the object whose 
behavior is to be predicted as a rational agent; then you figure out what 
beliefs that agent ought to have, given its place in the world and its purpose. 
Then you figure out what desires it ought to have, on the same considerations, 
and finally you predict that this rational agent will act to further its goals 
in the light of its beliefs. A little practical reasoning from the chosen set 
of beliefs and desires will in most instances yield a decision about what the 
agent ought to do; that is what you predict the agent will do.}
\end{quote} 

As Dennett argues, and it is obvious from our everyday application of the
intentional stance, this ``theory'' combines an amazing simplicity with a very
significant predictive power. Indeed, it is not evident at all how we can
(approximately) predict the behavior of systems as complex as a leopard or a
human using such a compact scheme. If we look at the physical and molecular
complexity of these systems from a detached point of view, it may seem that
saying \emph{anything whatsoever} about what they will do in the next seconds
is hopelessly impossible---and yet. The explanation of why the intentional
stance works is completely outside the scope of this note, but the truth is
that it works remarkably well. In fact, as we can also found in Dennett, it is
precisely this success of the theory what justifies our ontological commitment
to the actual existence of the entities that the theoretical terms ``belief'',
``desire'' or ``goal'' refer to. As \cite{Viger2000} clearly puts it:

\begin{quote}
{\small \ldots our ontological commitment to beliefs and desires derives from
the success of intentional explanation. There is simply no room, then, to 
contrast the existential status of entities posited by physics. Beliefs and
desires are as ontologically secure as electrons or gravitational fields, for 
it is nothing more than the predictive success of physics that grounds our
commitment to the entities that our physical theories posit.}
\end{quote}

\section{Conclusion}
\label{sec:conclusion}

To conclude, my notion of ontological commitment (which is very much aligned to
that of the mentioned philosophers and scientists, and possibly to that of many
others) can be briefly described as \emph{a personal working hypothesis about
what really exists out there, which is not arbitrary but partially validated
and justified by the success of the scientific theories that contain the
corresponding concepts or are suggested by them}.

I have also emphasized the influence of these hypotheses in what kind of
science one produces or the skill with which theories are understood and
applied, as well as the consequent importance of identifying, clarifying and
making explicit one's ontological commitments if fruitful scientific
discussions are to be had. This might seem obvious for the philosophically
inclined reader, but it is my experience that it is not quite so among
theoretical physicists.

\section*{Acknowledgements}

\hspace{0.5cm} I acknowledge with pleasure very useful conversations with Tim
Maudlin and Jean Bricmont. Not only their articles and books speak volumes, but
their willingness to answer silly questions by persistent amateurs is close to
legendary.

This work has been financially supported by the grant Grupo Consolidado 
``Biocomputaci\'on y F\'{\i}sica de Sistemas Complejos'' (DGA, Spain).

\phantomsection
\addcontentsline{toc}{section}{References}

\end{document}